\documentclass[twocolumn,showpacs,superscriptaddress,10pt]{revtex4-1} 
\usepackage{amsmath,amssymb,graphicx}
\begin{document}

\title{Nonlinear silicon photonics analyzed with the moment method}

\author{Simon Lefrancois}\email{Corresponding author: s.lefrancois@physics.usyd.edu.au}
\author{Chad Husko}
\affiliation{Centre for Ultrahigh bandwidth Devices for Optical Systems (CUDOS), Institute of Photonics and Optical Science (IPOS), School of Physics, University of Sydney, NSW 2006, Australia}
\author{Andrea Blanco-Redondo}
\affiliation{Centre for Ultrahigh bandwidth Devices for Optical Systems (CUDOS), Institute of Photonics and Optical Science (IPOS), School of Physics, University of Sydney, NSW 2006, Australia}
\affiliation{Aerospace Unit, Industry and Transport Division, Tecnalia, Parque Tecnologico de Bizkaia, Ibaizabal Bidea, Edificio 202 E-48170 Zamudio, Bizkaia, Spain}
\author{Benjamin J. Eggleton}
\affiliation{Centre for Ultrahigh bandwidth Devices for Optical Systems (CUDOS), Institute of Photonics and Optical Science (IPOS), School of Physics, University of Sydney, NSW 2006, Australia}

\begin{abstract}
We apply the moment method to nonlinear pulse propagation in silicon waveguides in the presence of two-photon absorption, free-carrier dispersion and free-carrier absorption. The evolution equations for pulse energy, temporal position, duration, frequency shift and chirp are obtained. We derive analytic expressions for the free-carrier induced blueshift and acceleration and show that they depend only on the pulse peak power. Importantly, these effects are independent of the temporal duration. The moment equations are then numerically solved to provide fast estimates of pulse evolution trends in silicon photonics waveguides.  We find that group-velocity and free-carrier dispersion dominate the pulse dynamics in photonic crystal waveguides. In contrast, two-photon and free-carrier absorption dominate the temporal dynamics in silicon nanowires. To our knowledge, this is the first time the moment method is used to provide a concise picture of multiphoton and free-carrier effects in silicon photonics. The treatment and conclusions apply to any semiconductor waveguide dominated by two-photon absorption.
\end{abstract}


\maketitle 

\section{Introduction}

Semiconductor waveguides offer the potential for highly integrated and low-power optical devices for applications such as optical communications~\cite{Kuo:06,Lee:09}, ultrafast measurements~\cite{Foster2008} and quantum communications~\cite{Xiong:11}. Silicon in particular combines very high nonlinearity~\cite{Leuthold2010} and well-developed micro-fabrication techniques~\cite{Lipson:2005}. In addition to displaying nonlinear phenomena common to all Kerr materials such as four-wave mixing~\cite{Fukuda:05} and soliton compression~\cite{ColmanP.2010}, silicon and other semiconductors exhibit multi-photon absorption and free-carrier effects. This gives rise to additional phenomena such as nonlinear losses~\cite{Yin:07}, the free-carrier induced blueshift~\cite{Monat:09} and pulse acceleration~\cite{Blanco-Redondo2014}. Given the large body of experimental work in silicon photonics, a deep theoretical understanding of multiphoton and free-carrier effects on optical pulse propagation is important to evaluate their impact on silicon and semiconductor photonics devices. This can also elucidate novel nonlinear wave phenomena.

Several theoretical approaches have been applied to silicon photonics devices. In general, the nonlinear equations for wave propagation can be solved numerically~\cite{Yin:07}. Alternatively, analytical methods provide a more fundamental understanding by exposing the phenomena underlying the observed effects. In some cases, the equations can be solved directly under some assumptions. For example, an integral solution for pulse propagation in silicon was obtained in the case of small variation of the temporal pulse~\cite{Rukhlenko:2010}. Such methods provide information on intrapulse temporal dynamics, but basic physical interactions are often embedded in implicit equations making it challenging to derive intuitive understanding. Alternatively, to elucidate the physical scalings of pulse propagation dynamics explicitly, we can derive a set of evolution equations for discrete parameters such as energy and pulse duration. These latter methods allow trends to be isolated and quickly evaluated over a broad range of parameters. A well-known example is the calculation of the Raman-induced soliton self-frequency shift in optical fibers~\cite{Gordon:86ssfs}. Recently, perturbation theory has been applied to soliton interaction with free-carriers effects~\cite{Saleh:2011,Roy:13}. This technique can directly estimate the soliton blueshift, but pulse parameters are limited to those of a fundamental soliton.

A more general analytic treatment of pulse propagation in silicon photonics should allow unconstrained parameters. The method of moments provides general trends by allowing arbitrary pulse shapes and parameter sets. The root mean square (RMS) quantities of interest for the pulse are defined and differential evolution equations are extracted~\cite{Perez:07}. The method of moments has been applied to glass media such as optical fibers to estimate jitter in optical communications~\cite{Gordon:86noise}, as well as model intrapulse Raman scattering~\cite{Santhanam2003413,Chen:10} and the nonlinear dynamics of short pulses~\cite{Tsoy:06,Burgoyne:07}. However, to our knowledge, the moment method has not been applied to semiconductor effects such as nonlinear losses and free-carrier interactions.

In this article, we apply the method of moments to pulse propagation in silicon photonics. We derive general equations for pulse evolution taking into account two-photon absorption (TPA) and free-carrier dispersion (FCD) and absorption (FCA) in addition to dispersion and self-phase modulation (SPM). Analytical expressions for the free-carrier induced blueshift and acceleration are obtained. These are found to depend explicitly on the pulse peak power only, due to the accumulated nature of free-carrier effects. Surprisingly, they are found to be independent of the pulse temporal duration. The moment equations are then solved numerically to study the influence of free-carrier effects in silicon photonic crystal and nanowire waveguides. Finally, the moment trends are compared with full numerical simulations and experiments, and good agreement is found over a broad range of input powers. The model developed here provides a concise and general picture of the complex pulse shaping effects observed in silicon waveguides directly in terms of the basic physical mechanisms underpinning them.

\section{The moment method in silicon waveguides}

Pulse propagation in semiconductor waveguides can be modelled using a generalized nonlinear Schr\"odinger equation (GNLSE)~\cite{Yin:07}. For semiconductor waveguides such as silicon where two-photon absorption is the dominant nonlinear absorption mechanism, the equation is:
\begin{multline}
\frac{\partial A}{\partial z} = -\frac{\alpha}{2}A -i\frac{\beta_2}{2}\frac{\partial^2 A}{\partial t^2}  +\frac{\beta_3}{6}\frac{\partial^3 A}{\partial t^3} +i\gamma|A|^2A\\ - \frac{\gamma_{TPA}}{2}|A|^2A+\bigg(in_{FC}k_0-\frac{\sigma}{2}\bigg)N_cA.
\label{eqn:nlse}
\end{multline}
Here $A(z,t)$ is the amplitude envelope (the power $P=|A|^2$), $N_c(z,t)$ is the free-carrier density, $\alpha$ is the linear propagation loss and $\beta_2$ and $\beta_3$ are the group-velocity dispersion (GVD) and third-order dispersion (TOD) coefficients, respectively. The Kerr nonlinear parameter causing self-phase modulation is $\gamma=n_2k_0/A_{eff}$ and $\gamma_{TPA}=\alpha_{TPA}/A_{eff}$ is the two-photon absorption parameter. The free-carrier absorption parameter is $\sigma$ and the carrier density-dependent refractive index coefficient $n_{FC}$ causes free-carrier dispersion. Note that $n_{FC}$ is generally negative since in a free-carrier plasma higher density lowers the refractive index~\cite{soref:87}. The bulk nonlinear parameters are the intensity dependent refractive index $n_2$ and absorption $\alpha_{TPA}$. The effective mode area is $A_{eff}$ and the vacuum wavevector is $k_0$.

For the free-carrier density in silicon, we consider that two-photon absorption is the dominant generation mechanism. This is the case when doping and carrier injection are absent. The free-carrier density is then governed by the equation:
\begin{equation}
\frac{\partial N_c}{\partial t}  = \rho_{FC}|A|^4-\frac{N_c}{\tau_c}.
\label{eqn:Nc}
\end{equation}
We have defined the free-carrier generation rate $\rho_{FC}=\alpha_{TPA}/(2\hbar\omega_0{A_{eff}}^2)$ with $\hbar\omega_0$ the photon energy, and $\tau_c$ is the free-carrier lifetime. From here on we will assume that the free-carrier recombination term is negligible. This requires $\tau_c$ to be much longer than an individual pulse but much shorter that the pulse train period. For silicon waveguides we have $\tau_c\sim1-10$~ns~\cite{Boyraz:04,Corcoran:10}, so this assumption is valid for femtosecond and picosecond pulses with megahertz repetition rates.

The method of moments reduces the dynamics of pulse propagation down to a set of discrete parameters. We start by defining the moments of energy $E$, acceleration $T_c$, spectral shift $\Omega$, RMS temporal width $\sigma_t$ and RMS chirp $\tilde{C}$~\cite{Santhanam2003413}:
\begin{align}
&E = \int_{-\infty}^{\infty}|A|^2dt, \label{eqn:E}\\
&T_c = \frac{1}{E}\int_{-\infty}^{\infty}t|A|^2dt, \\
&{\sigma_t}^2 = \frac{1}{E}\int_{-\infty}^{\infty}(t-T_c)^2|A|^2dt, \\
&\Omega = \frac{i}{2E}\int_{-\infty}^{\infty}(A^*\partial_tA-A\partial_tA^*)dt,\\
&\tilde{C} = \frac{i}{2E}\int_{-\infty}^{\infty}(t-T_c)(A^*\partial_tA-A\partial_tA^*)dt. \label{eqn:C}
\end{align}

Propagation equations for each moment can be obtained by taking the derivative of Eqs.~\ref{eqn:E}-\ref{eqn:C}, combining with the GNLSE in Eq.\ref{eqn:nlse} and using integration by parts, assuming the pulse amplitude vanishes at infinity~\cite{Perez:07}. To simplify the equations, we define the nonlinear loss coefficient:
\begin{equation}
\Gamma_{NL} = \gamma_{TPA}|A|^2 + \sigma N_c.
\end{equation}

We will now show the full, general moment equations. In practice not all terms contribute significantly given a set of waveguide and input pulse parameters, but for now we present the most general case. The full moment evolution equations are:
\begin{widetext}
\begin{align}
&\frac{dE}{dz} = -\alpha E - \int_{-\infty}^{\infty}\Gamma_{NL}|A|^2dt, \label{eqn:dEdz}\\
&\frac{dT_c}{dz} =  \beta_2\Omega + \frac{\beta_3}{2E}\int_{-\infty}^{\infty}|\partial_tA|^2dt 
 -\frac{1}{E}\int_{-\infty}^{\infty}(t-T_c)\Gamma_{NL}|A|^2dt, \\
&\frac{d\sigma_t}{dz} = \frac{\beta_2\tilde{C}}{\sigma_t} + \frac{\beta_3}{2\sigma_t E}\int_{-\infty}^{\infty}(t-T_c)|\partial_tA|^2dt 
 -\frac{1}{2E\sigma_t}\int_{-\infty}^{\infty}\big((t-T_c)^2-{\sigma_t}^2\big)\Gamma_{NL}|A|^2dt, \\
&\frac{d\Omega}{dz} = \frac{-n_{FC}k_0}{E}\int_{-\infty}^{\infty}|A|^2\partial_tN_cdt 
 +\frac{1}{E}\int_{-\infty}^{\infty}\bigg(\frac{A^*\partial_tA-A\partial_tA^*}{2i}+\Omega|A|^2\bigg)\Gamma_{NL}dt,  \\
&\frac{d\tilde{C}}{dz} = \frac{\beta_2}{E}\int_{-\infty}^{\infty}|\partial_t A|^2dt + i\frac{\beta_3}{4E}\int_{-\infty}^{\infty}(\partial_t^2A\partial_tA^*-\partial_t^2A^*\partial_tA)dt - \Omega\frac{dT_c}{dz} 
+ \frac{\gamma}{2E}\int_{-\infty}^{\infty}|A|^4dt \nonumber\\
&\hspace{3cm}  +\frac{-n_{FC}k_0}{E}\int_{-\infty}^{\infty}(t-T_c)|A|^2\partial_tN_cdt  + \frac{1}{E}\int_{-\infty}^{\infty}\bigg((t-T_c)\frac{A^*\partial_tA-A\partial_tA^*}{2i}+\tilde{C}|A|^2\bigg)\Gamma_{NL}dt. \label{eqn:dCdz}
\end{align}
\end{widetext}

\section{Moments of a hyperbolic secant pulse}

The moment Eqs.~\ref{eqn:dEdz}-\ref{eqn:dCdz} contain the full GNLSE dynamics without assumptions, but it is difficult to extract trends in this form. They can be reduced to ordinary coupled differential equations by specifying a pulse shape. The general shape is assumed to be maintained through propagation, while its parameters are allowed to vary continuously. Here we will use a chirped hyperbolic secant pulse, which often occurs in soliton-like propagation:
\begin{multline}
A(z,t) = \sqrt{\frac{E}{2\tau}}\text{sech}\bigg( \frac{t-T_c}{\tau} \bigg)\times\\
\exp\bigg(-i\Omega(t-T_c) -i\frac{C}{2{\tau}^2}(t-T_c)^2\bigg)
\label{eqn:sech}
\end{multline}

The pulse duration $\tau$ and chirp $C$ in Eq.~\ref{eqn:sech} are linked to the RMS values by the constant $K=12/\pi^2\approx1.22$ as $\tau^2=K{\sigma_t}^2$ and $C=K\tilde{C}$~\cite{Santhanam2003413}. By combining Eqs.~\ref{eqn:dEdz}-\ref{eqn:dCdz} with Eq.~\ref{eqn:sech}, we obtain the set of moment equations:
\begin{align}
\frac{dE}{dz} &= -\alpha E  - \frac{\gamma_{TPA}}{3}\frac{E^2}{\tau} - \frac{\sigma\rho_{FC}}{6}\frac{E^3}{\tau} \label{eqn:dEdzsech}, \\
\frac{dT_c}{dz} &= \beta_2\Omega + \frac{\beta_3}{2}\bigg( \Omega^2 + \frac{1+\frac{\pi^2}{4}C^2}{3\tau^2} \bigg) -\frac{7\sigma\rho_{FC}}{72}E^2, \label{eqn:dTdzsech}\\
\frac{d\tau}{dz} &= (\beta_2+\beta_3\Omega)\frac{C}{\tau} + \frac{\gamma_{TPA}}{\pi^2}E, \label{eqn:dtaudzsech}\\
\frac{d\Omega}{dz} &= -\frac{2}{15}n_{FC}k_0\rho_{FC}\frac{E^2}{\tau^2} -\frac{7}{72}\sigma\rho_{FC}\frac{C E^2}{\tau^2}, \label{eqn:dWdzsech}\\
\frac{dC}{dz} &= (\beta_2+\beta_3\Omega) \frac{\frac{4}{\pi^2} + C^2}{\tau^2} + \frac{2}{\pi^2}(\gamma+\gamma_{TPA}C)\frac{E}{\tau}. \label{eqn:dCdzsech}
\end{align}
Notice that the pulse duration Eq.~\ref{eqn:dtaudzsech} and chirp Eq.~\ref{eqn:dCdzsech} depend on the effective GVD $\beta_2' = \beta_2+\beta_3\Omega$ experienced by a frequency-shifted pulse~\cite{Tsoy:06}. The current equations do not account for pure TOD broadening, usually a small effect for pulses longer than 100 fs. Including this would require a pulse ansatz allowing asymmetry and higher order phase, which would complicate the equations.

These results are consistent with those from soliton perturbation theory including TPA and FCD~\cite{Roy:13}, noting that for a secant pulse the peak power $P = E/(2\tau)$. The moment method employed in this work is more general as it allows pulse duration and chirp to vary freely. Consequently, the moment method has a broader range of applicability across multiple pulse shapes beyond the assumptions underlying perturbation theory.

In addition to the well-known dispersion and SPM effects~\cite{Burgoyne:07}, the system of moments in Eqs.~\ref{eqn:dEdzsech}-\ref{eqn:dCdzsech} capture several important multiphoton and free-carrier effects unique to semiconductor waveguides. These are summarised in Table~\ref{tab:effects}. 
\begin{table}[h!]
  \caption{Multi-photon and free-carrier effects in silicon.}
  \label{tab:effects}
  \begin{center}
    \begin{tabular}{lc}
    \hline
Two-photon absorption & $\gamma_{TPA}E^2/\tau$ \\
Free-carrier absorption & $\sigma\rho_{FC}E^3/\tau$ \\
FCD acceleration & $\beta_2\Omega_{FCD}$\\
FCA pulse trail suppression & $\sigma\rho_{FC}E^2$ \\ 
TPA effective broadening & $\gamma_{TPA}E$ \\
FCD blueshift & $|n_{FC}|k_0\rho_{FC}E^2/\tau^2$ \\
FCA chirped frequency shift & $-\sigma\rho_{FC}C E^2/\tau^2$ \\
TPA chirp redistribution & $\gamma_{TPA}CE/\tau$ \\
    \hline
    \end{tabular}
  \end{center}
\end{table}

The energy Eq.~\ref{eqn:dEdzsech} accounts for nonlinear loss due to two-photon and free-carrier absorption. It is also clear that the FCA effect is second-order to the dominant TPA effect from their respective cubic and square energy dependence. The FCA term in the temporal shift Eq.~\ref{eqn:dTdzsech} is caused by accumulated free-carriers suppressing the trailing edge of the pulse, shifting its center of mass toward the leading edge. This is usually weak in PhC waveguides compared to the dispersive effects, which shift the entire pulse~\cite{Blanco:optica14}. The TPA term in the duration Eq.~\ref{eqn:dtaudzsech} represents effective pulse broadening due to TPA suppressing the pulse peak more strongly and is also second-order to the dispersive effect in PhC. However, the TPA and FCA temporal effects can be dominant in nanowire waveguides due to their smaller dispersion~\cite{Liao:13}. The FCD blueshift can be seen in Eq.\ref{eqn:dWdzsech} (remember that $n_{FC}<0$). When coupled with the dispersion term in Eq.~\ref{eqn:dTdzsech}, the blueshift causes pulse acceleration or deceleration, depending on the sign of $\beta_2$.

The FCA frequency shift in Eq.\ref{eqn:dWdzsech} is caused by asymmetric absorption of a chirped pulse and is usually small. For example, using the parameters in section~\ref{sec:propsilicon} below, when $C<|n_{FC}|k_0/\sigma=15$, the FCA shift is small compared to the FCD blueshift. Similarly, the TPA term in the chirp Eq.~\ref{eqn:dCdzsech} is due to power-dependent suppression of a chirped pulse and is also often small. For the parameters below, the SPM chirp is dominant for $C < \gamma/\gamma_{TPA} = 2.4$. The FCD contribution to chirp in Eq.~\ref{eqn:dCdz} cancels out for symmetric pulses, similar to Raman scattering~\cite{Santhanam2003413}.

Another common pulse shape is the chirped gaussian. The resulting moment equations are similar to Eqs.~\ref{eqn:dEdzsech}-\ref{eqn:dCdzsech} with different numerical factors and are given in Appendix~\ref{app:gauss}. The essential physics remain the same.

\section{Free-carrier induced blueshift and acceleration}
\label{sec:blueshift}

We now use the moment equations to derive new physical insight about the free-carrier induced frequency and temporal shifts. The first term in Eq.~\ref{eqn:dWdzsech} represents the free-carrier induced blueshift. This general feature of free-carrier dispersion in plasmas has been predicted and observed in gases~\cite{Wood:1991}, silicon waveguides~\cite{Monat:09,Roy:13,Blanco-Redondo2014} and gas-filled hollow-core fibers~\cite{Fedotov:2007,Saleh:2011}. As discussed above, the second term is due to FCA on chirped pulses, which is usually smaller than the FCD term for small chirps and therefore negligible. 
 
The $E^2/\tau^2$ dependence of the free-carrier blueshift corresponds to a $P^2$ dependance. The pulse duration dependence of the FCD phase cancels out because free-carriers accumulate over the pulse ($N_c$ is the integral of Eq.~\ref{eqn:Nc}). Thus, at constant peak power the free-carrier blueshift is not affected by pulse duration. This is a key conclusion of this work. Note that a $P^3$ dependence is expected of a three-photon limited semiconductor system~\cite{ColmanP.2010}, while for near-threshold tunnelling ionization in gases this would scale as $P$~\cite{Saleh:2011}. As an experimental consideration, we note other effects which suppress peak power, such as dispersion, become weaker for longer pulses. Consequently, the observed blueshift may be larger for longer pulses in these systems.

The pulse duration independence of the FCD blueshift contrasts with the well studied Raman self-frequency shift (SFS). In general, Raman SFS scales like $E/\tau^3$~\cite{Santhanam2003413}, corresponding to a $P/\tau^2$ dependence. This is expected since Raman scattering depends on the local pulse steepness (derivative in \cite[Eq. 1]{Santhanam2003413}) which increases for shorter pulses. 

We now estimate the blueshift accumulated during propagation. Defining $P(z)= E/(2\tau) =P_0\eta(z)$, with the input peak power $P_0$ and peak power decay factor $\eta \le 1$, and assuming $n_{FC}<0$, the blueshift becomes:
\begin{align}
\Omega(z) &= \frac{8}{15}|n_{FC}|k_0\rho_{FC}\int_0^zP(z')^2dz', \nonumber\\
&= \frac{8}{15}|n_{FC}|k_0\rho_{FC}{P_0}^2z_{eff}.
\label{eqn:blueshift}
\end{align}

We have defined the FCD effective length:
\begin{equation}
z_{eff} = \int_0^z\eta(z')^2dz'.
\end{equation}
The effective length $z_{eff}$ capture the effects of linear and nonlinear losses and other mechanisms reducing peak power, such as dispersion. For long propagation, $z_{eff}$ converges to a characteristic length $L_{max}$ discussed below. We can also show that for short propagation $z\ll L_{max}$ we have $z_{eff}\approx z$. Thus the blueshift is initially linear in $z$ before converging to a maximum value.

We now estimate the free-carrier induced temporal shift. To first order, consider the GVD term in Eq.~\ref{eqn:dTdzsech}:
\begin{align}
T_c(z) &=  \frac{8}{15}\beta_2|n_{FC}|k_0\rho_{FC}{P_0}^2\int_0^zz_{eff}'dz'.
\label{eqn:dtzeff}
\end{align}
For short propagation the temporal shift $T_c \propto z^2$, corresponding to a constant free-carrier induced pulse acceleration. When the blueshift saturates we have $T_c \propto z$, a fixed velocity from the accumulated free-carrier blueshift.

It is often possible to obtain a closed expression for the effective length $z_{eff}$. Moreover, this effective length converges to a saturated non-linear length $L_{max}$ after sufficient propagation. Table~\ref{tab:zeff} summarizes the effective lengths for the main peak power decay mechanisms in silicon photonics. The decay factors for TPA and linear loss can obtained from the NLSE Eq.~\ref{eqn:nlse}. The peak power reduction from dispersive broadening is obtained by solving Eq.~\ref{eqn:dtaudzsech} and Eq.~\ref{eqn:dCdzsech} together with GVD only and zero input chirp. The dispersion length is $L_D = \tau(0)^2/|\beta_2|$.

\begin{table}[h!]
  \caption{Effective lengths for FCD effects.}
  \label{tab:zeff}
  \begin{center}
    \begin{tabular}{|l|c|c|c|}
    \hline
  & $\eta(z)$ & $z_{eff}$ & $L_{max}$ \\
      \hline
TPA & $(1+\gamma_{TPA}P_0z)^{-1}$ & $z/(1+\gamma_{TPA}P_0z)$ & $1/(\gamma_{TPA}P_0)$ \\      \hline
Loss &  $e^{-\alpha z}$ & $(1-e^{-2\alpha z})/2\alpha$& $1/(2\alpha)$ \\      \hline
GVD & $1/\sqrt{1+\frac{4}{\pi^2}\frac{z^2}{{L_D}^2}}$  & $\frac{\pi}{2}L_D\tan^{-1}{\bigg( \frac{2}{\pi}\frac{z}{L_D} \bigg)}$ & $(\pi/2)^2L_D$\\  
    \hline
    \end{tabular}
  \end{center}
\end{table}

\section{Pulse propagation in silicon PhC waveguides}
\label{sec:propsilicon}

In this section we study the propagation of pulses in silicon waveguides and identify the dominant contributions to the nonlinear pulse dynamics. The system of moment Eqs.~\ref{eqn:dEdzsech}-\ref{eqn:dCdzsech} can be solved numerically in the general case. Solving these simple equations is much faster than full numerical solutions of the NLSE and can quickly yield trends for pulse propagation in semiconductors, as long as the pulse shape does not deviate strongly from the input. The computational advantage is particularly strong here since full simulations require the NLSE and free-carrier equations to be solved together.

The parameters used correspond to a slow-light silicon photonic crystal (PhC) waveguide described elsewhere~\cite{Blanco-Redondo2014,Blanco:optica14}. At a wavelength $\lambda_0$ of 1543~nm, the effective modelling parameters are given in Table~\ref{tab:param}. The slow-light scaling of the bulk parameters used to obtain these is discussed in Appendix~\ref{app:slowlight}.
\begin{table}[h!]
  \caption{Physical parameters for a silicon PhC waveguide at 1543~nm}
  \label{tab:param}
  \begin{center}
    \begin{tabular}{lclc}
    \hline
	loss & 5.0 dB/mm & $L$ & 0.396 mm \\
	$\beta_2$ & -10.4 ps$^2$/mm & $\beta_3$ & -0.53 ps$^3$/mm\\
	$\gamma$ & 2948 W$^{-1}$m$^{-1}$ & $\gamma_{TPA}$ & 1207 W$^{-1}$m$^{-1}$\\
	$n_{FC}$ & -3.4$\cdot 10^{-26}$ m$^2$ & $\sigma$ & 9.0$\cdot10^{-21}$ m$^2$ \\
	$\rho_{FC}$ & $1.46\cdot10^{34}$ m$^{-3}$W$^{-2}$s$^{-1}$ &\\
    \hline
    \end{tabular}
  \end{center}
\end{table}

The input pulse parameters are $E(0) = 10.4$~pJ, $\tau(0) = 0.74$~ps and $C(0) = 0$, corresponding to an unchirped hyperbolic secant with a full-width half-maximum (FWHM) duration of 1.3~ps and 4~W peak power. The coupled moment Eqs.~\ref{eqn:dEdzsech}-\ref{eqn:dCdzsech} are solved using a 4th-order Runge-Kutta integrator. We present results with different effects turned off to isolate their influence. The calculations without TPA correspond to dropping the power-dependent loss from Eq.~\ref{eqn:nlse} while keeping free-carrier generation active. Although this is not possible in reality, it allows the influence of multiphoton loss to be isolated from that of free-carrier effects.

\begin{figure}[htbp]
\centerline{\includegraphics[width=3in]{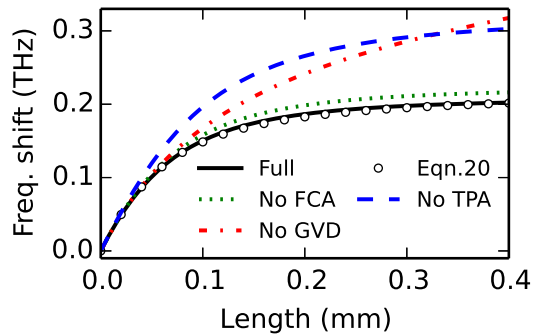}}
\caption{Frequency shift $\Delta f = \Omega/(2\pi)$ of a hyperbolic secant pulse along a silicon PhC waveguide from the full moment equations with different effects suppressed (lines). Also shown is the analytic result from Eq.\ref{eqn:blueshift} including FCD blueshift only (circles).}
\label{fig:momentdf}
\end{figure}

The free-carrier effects are most pronounced when looking at the frequency and temporal shifts. The blueshift $\Omega$ is caused by free-carrier dispersion, and free-carrier absorption has a minor effect here. The frequency shift $\Delta f = \Omega/(2\pi)$ is shown in Fig.~\ref{fig:momentdf}. The blueshift accumulates mostly over one effective length. The lengths scales as defined in section~\ref{sec:blueshift} are $L_{max}^{D} = 0.13$~mm, $L_{max}^{TPA} = 0.21$~mm and $L_{max}^{lin} = 0.44$~mm. In this case, the blueshift is limited mostly by dispersion and two-photon absorption, while linear loss plays a minor role. With TPA loss removed a larger blueshift is accumulated until dispersion limits the FCD again. For comparison, the blueshift predicted by the approximate integral Eq.~\ref{eqn:blueshift} is also plotted. To account for the contribution of different effects to $z_{eff}$, we apply the effective lengths in cascade, from the strongest to the weakest effect. Thus, $z_{eff}^{D}$ is first computed and used as the input for $z_{eff}^{TPA}$. There is close agreement with the full moments solution, so Eq.~\ref{eqn:blueshift} can provide a good estimate of the blueshift without having to calculate all the moments.

\begin{figure}[htbp]
\centerline{\includegraphics[width=3in]{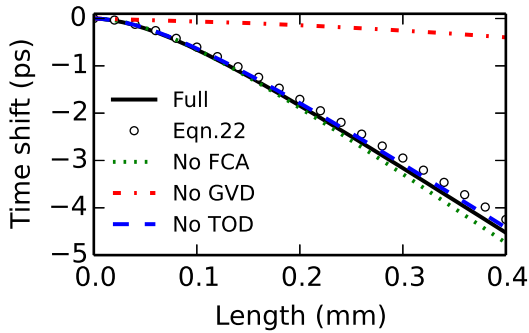}}
\caption{Temporal shift $T_c$ of a hyperbolic secant pulse along a silicon PhC waveguide from the full moment equation with different effects suppressed (lines). Also shown is the analytic result from Eq.\ref{eqn:dtzeff} with GVD only (circles).}
\label{fig:momentTc}
\end{figure}

The temporal shift $T_c$ is shown in Fig.~\ref{fig:momentTc}. When the free-carrier blueshift is combined with anomalous dispersion, the pulse advances in time compared to the linear case~\cite{Blanco-Redondo2014}. At normal dispersion the pulse would lag instead. The relatively small negative TOD adds a minor contribution here. In contrast to the frequency shift, the temporal shift occurs throughout the whole waveguide since it depends on the accumulated blueshift, not the local peak power. The temporal shift $T_c$ increases quadratically in $z$ as long as the blueshift is increasing linearly in $z$, as described by Eq.~\ref{eqn:dtzeff}. However, as the peak power decreases during propagation due to dispersion and loss during propagation, the FCD blueshift saturates and $z$ approaches a constant $L_{max}$ and $T_c$ increases linearly, as expected. For comparison, results from the effective length treatment from Eq.~\ref{eqn:dtzeff} are shown, where the $z_{eff}$ used in the blueshift calculation was integrated numerically. The trend is well accounted for, but the temporal shift is slightly underestimated since TOD and FCA are not included in the approximation.

\begin{figure}[htbp]
\centerline{\includegraphics[width=3in]{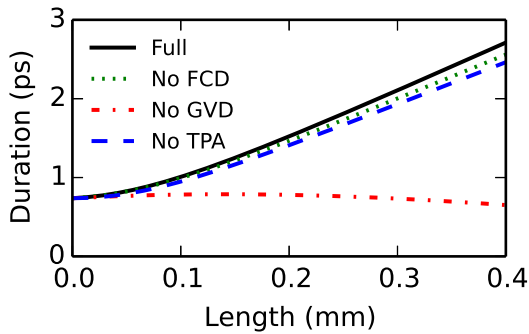}}
\caption{Pulse duration $\tau$ of a hyperbolic secant pulse along a silicon PhC waveguide with different effects suppressed.}
\label{fig:momenttau}
\end{figure}

The evolution of the pulse duration $\tau$ is shown in Fig.~\ref{fig:momenttau}. This is dominated by dispersive broadening, while the free-carrier effects have little incidence at this input energy. Pulse broadening is slightly reduced with TPA suppressed. TPA slightly broaden the pulse as in Eq.~\ref{eqn:dtaudzsech}, and this is enhanced by the nonlinear loss which reduces the SPM pulse compression for anomalous dispersion. FCD contributes only slightly to temporal broadening. Removing TOD (not shown) has the exact same effect. This is because the blueshift increases the effective GVD $\beta_2' = \beta_2+\beta_3\Omega$ at the pulse center frequency when $\beta_2$ and $\beta_3$ have the same sign, so this is a combined FCD-TOD effect~\cite{Blanco:optica14}. The trend for the chirp moment is similar to the one for pulse duration and is not shown here.

\begin{figure}[htbp]
\centerline{\includegraphics[width=3in]{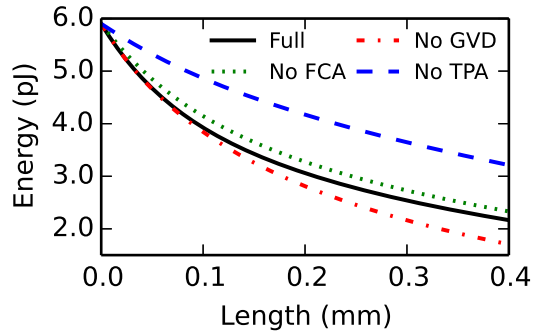}}
\caption{Energy $E$ of a hyperbolic secant pulse in a silicon PhC waveguide with different effects suppressed.}
\label{fig:momentE}
\end{figure}

The evolution of the pulse energy $E$ is shown in Fig.~\ref{fig:momentE}. Two-photon absorption is the dominant loss mechanism over the characteristic length $L_{max}^{TPA}$, and linear loss also contributes over the whole length. Free-carrier absorption plays a minor role here. Removing dispersion enhances nonlinear losses slightly since higher peak powers are maintained without dispersive broadening.

Overall, the FCD blueshift in Eq.~\ref{eqn:dWdzsech} and TPA loss in Eq.~\ref{eqn:dEdzsech} are clearly the dominant terms, as mentioned before. Thus, in many cases it is sufficient to consider only these two when solving Eq.~\ref{eqn:dEdzsech}-\ref{eqn:dCdzsech} to obtain the main trends.

\section{Comparison with simulations and experiments in silicon PhC waveguides}

We now compare the predictions of the moment method with experiments and full numerical simulations of the GNLSE given by Eq. \ref{eqn:nlse}. The experiments used a high sensitivity frequency-resolved electrical gating (FREG) system to probe the temporal intensity and phase of the output pulses~\cite{Dorrer:02}. The waveguide is a silicon PhC with parameters given in Table~\ref{tab:param}, and the details of the experiments are given in~\cite{Blanco-Redondo2014,Blanco:optica14}. The input pulses had a fixed input duration of about 1.3~ps FWHM ($\tau=0.74$~ps RMS) and the input peak power inside the waveguide was varied over several watts. The numerical simulations used the split-step Fourier method to solve Eqs.~\ref{eqn:nlse}-\ref{eqn:Nc} with the same device parameters and input pulse as the moment method.

\begin{figure}[htbp]
\centerline{\includegraphics[width=3.25in]{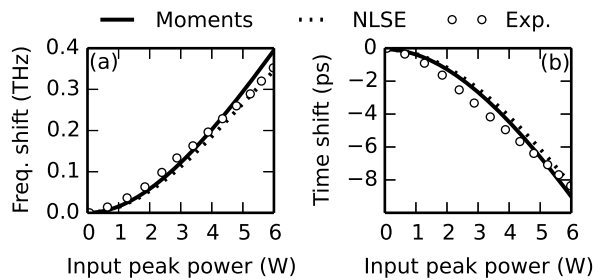}}
\caption{(a) Frequency shift $\Delta f = \Omega/(2\pi)$ and (b) temporal shift $T_c$ from the moment equations, NLSE simulations and experimental measurements for a silicon PhC waveguide at increasing input peak power.}
\label{fig:expdf}
\end{figure}

We first consider the free-carrier blueshift $\Omega$ in Fig.~\ref{fig:expdf}(a). The blueshift increases quadratically with input peak power over most of the range, as expected from Eq.~\ref{eqn:blueshift}. The moment method agrees well with NLSE simulations and experiments, while slightly overestimating at high powers. This is because the moment method underestimates pulse broadening, which will be discussed below in relation to the pulse duration. The trends for the temporal shift $T_c$ are shown in Fig.~\ref{fig:expdf}(b). Free-carrier acceleration is clearly observed, and agreement is fairly consistent as it was for the blueshift.

\begin{figure}[htbp]
\centerline{\includegraphics[width=3.25in]{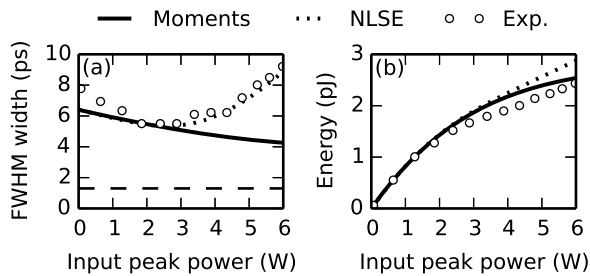}}
\caption{(a) Effective FWHM duration from the moment equations ($1.763\tau$) and NLSE simulations ($1.763\cdot1.1\cdot\sigma_t$), and experimental FWHM duration for a silicon PhC waveguide at increasing input peak power. The dashed line indicates the input pulse duration. (b) Pulse energy $E$ from the moment equations, NLSE simulations and experimental measurements.}
\label{fig:exptau}
\end{figure}

The trends for pulse duration are shown in Fig.~\ref{fig:exptau}(a). To better compare with experiments, we use the FWHM duration which is less sensitive to experimental noise. The effective FWHM for the moment method and NLSE are obtained by scaling the RMS duration by the shape factor ($t_{FWHM}=2\cosh^{-1}(\sqrt{2})\sqrt{12/\pi^2}\sigma_t$ for hyperbolic secant). The agreement is quite good up to 3~W. Dispersive broadening dominates while SPM partially cancels dispersion as power increases. For high powers the moment method and simulations diverge considerably. As we discuss in detail in a recent submission~\cite{Blanco:optica14}, this is caused by free-carrier dispersion induced spectral broadening, which induces temporal broadening when coupled with dispersion in the NLSE simulations. The moment method assumes a fixed pulse shape, so it instead predicts further SPM compression. As discussed before, the trend for chirp is similar to that for pulse duration and is not shown.

However, as shown above, predictions for the other moments remain accurate at high powers. The moment method's inability to predict FCD temporal broadening does not affect other predictions as much, since the nonlinear effects like blueshift and TPA accumulate mostly in the first part of propagation before dispersion and losses suppress them.
In contrast, most of the dispersive broadening occurs after significant nonlinear effects and losses have accumulated, explaining why discrepancies are stronger for pulse duration predictions.

Finally, results for pulse energy $E$ are shown in Fig.~\ref{fig:exptau}(b). The general trends agree well, with TPA causing significant nonlinear loss. There is some discrepancy at high powers which could be due to uncertainties in the effective area and nonlinear coefficients. 

\section{Temporal dynamics in silicon nanowires}

We now apply the method of moments to a silicon nanowire waveguide. The waveguide parameters provided by these devices will allows us to highlight effects that were not readily observable in the photonic crystal waveguides discussed so far. We will focus on effective pulse broadening caused by peak power suppression by two-photon absorption and temporal shifting due to the asymmetry of free-carrier absorption.

We use waveguide parameters similar to previous reports~\cite{Liao:13}. The input pulses have a FWHM duration of 1.3~ps ($\tau=0.74$~ps) and a centre wavelength of 1545~nm. The nanowire has a length $L=4$~mm, linear loss of 3~dB/cm and dispersion $\beta_2=-1.5$~ps$^2$/m. The effective non-linearity is $\gamma = 349$~W$^{-1}$m$^{-1}$ and the two-photon absorption is $\gamma_{TPA} = 143$~W$^{-1}$m$^{-1}$. The carrier generation rate is $\rho_{FC}=7.9\cdot10^{33}$ m$^{-3}$W$^{-2}$s$^{-1}$. The free-carrier dispersion coefficient is $n_{FC}= 1.35\cdot10^{-27}$~m$^3$ and the free-carrier absorption is $\sigma = 1.45\cdot10^{-21}$ m$^2$~\cite{Yin:07}.

\begin{figure}[htbp]
\centerline{\includegraphics[width=3in]{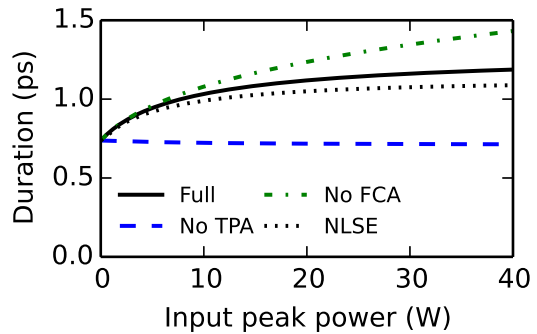}}
\caption{Pulse duration $\tau$ in a silicon nanowire at increasing input peak power from the moments method with different effects suppressed and compared to NLSE simulations.}
\label{fig:tau_nw}
\end{figure}

The output pulse duration from the moments method for increasing input peak power is shown in Fig.~\ref{fig:tau_nw} and compared to full NLSE simulations. Dispersive broadening is negligible in the nanowire ($L_D=363$~mm). Instead, two-photon absorption is the dominant broadening mechanism, as described by the energy-dependent term in Eqn.~\ref{eqn:dtaudzsech}. The TPA suppresses the peak of the pulse more strongly, resulting in an effective broadening. With FCA turned off, other losses are reduced and thus the TPA effects are strengthened. The quantitative deviation between moments and the NLSE are probably due to the pulse evolving away from the input shape.

\begin{figure}[htbp]
\centerline{\includegraphics[width=3in]{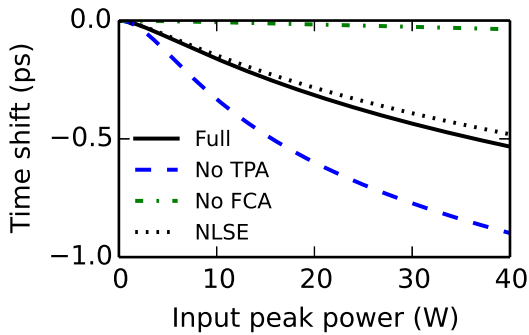}}
\caption{Temporal shift $T_c$ in a silicon nanowire at increasing input peak power from the moments method with different effects suppressed and compared to NLSE simulations.}
\label{fig:accell_nw}
\end{figure}

The output temporal shift for different input powers in the nanowire is shown in Fig.~\ref{fig:accell_nw}. This is dominated by free-carrier absorption, as described by the energy-dependent term in Eqn.~\ref{eqn:dTdzsech}. The accumulation of free-carriers towards the trailing edge of the pulse causes asymmetric absorption, causing the pulse temporal centroid to shift to earlier times. Removing TPA reduces losses and thus increases the free-carrier density and asymmetric absorption. In contrast, in the photonic crystal waveguides the temporal shift is caused by the free-carrier dispersion coupled with GVD.

Overall, the trends modelled for nanowires are similar to previous reports~\cite{Liao:13}. The silicon nanowires represent a distinct regime where dispersion is small and nonlinear absorption from multi-photon and free-carrier effects dominate temporal dynamics. In contrast, PhC devices are dominated by linear and free-carrier dispersions which redistribute energy rather than selectively suppress it. 

\section{Conclusion}

We have developed a moment method approach to free-carrier effects in semiconductors including two-photon absorption, free-carrier dispersion and free-carrier absorption. General expressions for the evolution of energy, temporal acceleration, pulse duration, frequency shift and chirp were obtained. The system of evolution equations for a hyperbolic secant pulse was obtained. From this we derived the free-carrier blueshift and temporal acceleration based on an effective length method accounting for linear and nonlinear peak power losses. We showed that the blueshift and acceleration depend explicitly on the pulse peak power only and are independent of the pulse duration, due to the buildup of free-carriers across the pulse.

We modelled pulse evolution in a silicon photonic crystal and nanowire waveguides and found that the free-carrier blueshift is limited mainly by dispersion and nonlinear absorption. The free-carrier absorption was found to play a negligible role in photonic crystals. In contrast, free-carrier and two-photon absorption are dominant in nanowires. Finally, comparing the moment method with full numerical simulations and experiments revealed good agreement for free-carrier phenomena such as multiphoton absorption, blueshifting and temporal acceleration. The temporal pulse width and chirp can be accurately predicted as long as free-carrier spectral broadening coupled with dispersion does not significantly distort the temporal pulse.

Overall, the moment method is a powerful tool for isolating the basic scaling laws behind complex multiphoton and free-carrier pulse shaping effects in silicon photonics. It also allows rapid calculation of pulse evolution trends in actual silicon waveguides to understand and optimise their design. The results and methods developed here apply to any two-photon absorption semiconductor waveguides and can be straightforwardly extended to n-photon absorption systems.

\appendix

\section{Moment equations for Gaussian pulses}
\label{app:gauss}

Here we derive the moment equations for the common case of a chirped gaussian pulse. We assume the pulse is given by:
\begin{multline}
A(z,t)= \sqrt{\frac{E}{\sqrt{\pi}\tau}}\exp\bigg(-\frac{1+iC}{2{\tau}^2}(t-T_c)^2 \\
-i\Omega(t-T_c) \bigg).
\label{eqn:gauss}
\end{multline}
Similarly to the hyperbolic secant above, the pulse duration and chirp parameters are linked to the RMS values by the constant $K=2$ as $\tau^2=K{\sigma_t}^2$ and $C=K\tilde{C}$~\cite{Santhanam2003413}.

By combining Eqs.~\ref{eqn:dEdz}-\ref{eqn:dCdz} with Eq.~\ref{eqn:gauss}, the moment equations are found to be:
\begin{align}
\frac{dE}{dz} &= -\alpha E - \frac{\gamma_{TPA}}{\sqrt{2\pi}}\frac{E^2}{\tau} - \frac{\sigma\rho_{FC}}{2\sqrt{2\pi}} \frac{E^3}{\tau}, \\
\frac{dT_c}{dz} &= \beta_2\Omega + \frac{\beta_3}{2}\bigg( \Omega^2 +\frac{1+C^2}{2\tau^2}\bigg) -\frac{\sigma\rho_{FC}}{2\sqrt{3}\pi}E^2, \\
\frac{d\tau}{dz} &= (\beta_2+\beta_3\Omega)\frac{C}{\tau} + \frac{\gamma_{TPA}}{4\sqrt{2\pi}}E, \\
\frac{d\Omega}{dz} &= \frac{-n_{FC}k_0\rho_{FC}}{\sqrt{3}\pi}\frac{E^2}{\tau^2} - \frac{\sigma\rho_{FC}}{2\sqrt{3}\pi}\frac{C E^2}{\tau^2}, \\
\frac{dC}{dz} &= (\beta_2+\beta_3\Omega)\frac{1+C^2}{\tau^2} + (\gamma+\frac{\gamma_{TPA}C}{2})\frac{E}{\sqrt{2\pi}\tau}.
\end{align}
The dependence on moments and physical parameters is the same as for the hyperbolic secant case, as expected since both pulse shape are symmetric and smoothly decaying. The quantitative factors are different, but of the same order. For instance, the FCD blueshift constant for a gaussian is $1/(\sqrt(3)\pi) \approx 0.18$, while for a hyperbolic secant it is $2/15 \approx 0.13$. If we convert from energy to peak power, this becomes $1/\sqrt{3}\approx0.58$ for a gaussian and $8/15\approx0.53$ for a secant, even closer.

\section{Slow-light scaling of bulk parameters in photonic crystal waveguides}
\label{app:slowlight}

Slow-light photonic crystal waveguides strongly enhance non-linear interactions by reducing the group-velocity of optical pulses~\cite{bhat:2001}. This can be included in the regular NLSE by scaling the bulk parameters by a slow-light factor $S=n_g/n_0$~\cite{Monat:09}.

The parameters in Table~\ref{tab:param} were obtained using $S=6.2$ from an experimentally measured $n_g=21.5$ combined with the refractive index of silicon $n_0=3.46$~\cite{Blanco:optica14}. The Kerr index is then $n_2=S^2\cdot6\cdot10^{-18}$ m$^2/W$ and the two-photon absorption is $\alpha_{TPA} = S^2\cdot1\cdot10^{-12}$ m/W~\cite{bristow:07}. The free-carrier absorption is $\sigma = S\cdot 1.45\cdot10^{-21}$ m$^2$~\cite{Yin:07}. For the free-carrier dispersion, we use $n_{FC} = S\cdot5.5\cdot10^{-27}$ m$^3$. This is higher than previously reported value, which could be due to the different scalings of the electron and hole contribution to free-carriers~\cite{soref:87}. The linear loss is $\alpha = S\cdot0.8$~dB/mm and the effective area is $A_{eff} = 0.32 \mu$m$^2$.

\begin{acknowledgments}

This work was supported by the Center of Excellence CUDOS (CE110001018) scheme of the Australian Research Council (ARC). C.H. was supported by the ARC Discovery Early Career Researcher award (DECRA-DE120102069). A.B.-R. was supported by the Tecnalia International Fellowship for Experienced Researchers and co-funded by Tecnalia Research and Innovation and the European Commission under the 7th Framework Programme (COFUND-People-Marie Curie Actions).

We thank Thomas F. Krauss and Juntao Li for providing the silicon photonic crystal waveguides.

\end{acknowledgments}


\begin{thebibliography}{10}
\newcommand{\enquote}[1]{``#1''}

\bibitem{Kuo:06}
Y.-H. Kuo, H.~Rong, V.~Sih, S.~Xu, M.~Paniccia, and O.~Cohen,
  \enquote{{Demonstration of wavelength conversion at 40 Gb/s data rate in
  silicon waveguides},} Opt. Express \textbf{14}, 11721--11726 (2006).

\bibitem{Lee:09}
B.~G. Lee, A.~Biberman, A.~C. Turner-Foster, M.~A. Foster, M.~Lipson, A.~L.
  Gaeta, and K.~Bergman, \enquote{{Demonstration of Broadband Wavelength
  Conversion at 40 Gb/s in Silicon Waveguides},} Photonics Technology Letters,
  IEEE \textbf{21}, 182--184 (2009).

\bibitem{Foster2008}
M.~A. Foster, R.~Salem, D.~F. Geraghty, A.~C. Turner-Foster, M.~Lipson, and
  A.~L. Gaeta, \enquote{{Silicon-chip-based ultrafast optical oscilloscope},}
  Nature \textbf{456}, 81--84 (2008).

\bibitem{Xiong:11}
C.~Xiong, C.~Monat, A.~S. Clark, C.~Grillet, G.~D. Marshall, M.~J. Steel,
  J.~Li, L.~O'Faolain, T.~F. Krauss, J.~G. Rarity, and B.~J. Eggleton,
  \enquote{{Slow-light enhanced correlated photon pair generation in a silicon
  photonic crystal waveguide},} Opt. Lett. \textbf{36}, 3413--3415 (2011).

\bibitem{Leuthold2010}
J.~Leuthold, C.~Koos, and W.~Freude, \enquote{{Nonlinear silicon photonics},}
  Nat Photon \textbf{4}, 535--544 (2010).

\bibitem{Lipson:2005}
M.~Lipson, \enquote{{Guiding, modulating, and emitting light on
  Silicon-challenges and opportunities},} Lightwave Technology, Journal of
  \textbf{23}, 4222--4238 (2005).

\bibitem{Fukuda:05}
H.~Fukuda, K.~Yamada, T.~Shoji, M.~Takahashi, T.~Tsuchizawa, T.~Watanabe, J.-i.
  Takahashi, and S.-i. Itabashi, \enquote{{Four-wave mixing in silicon wire
  waveguides},} Opt. Express \textbf{13}, 4629--4637 (2005).

\bibitem{ColmanP.2010}
P.~Colman, C.~Husko, S.~Combrie, I.~Sagnes, C.~W. Wong, and A.~{De Rossi},
  \enquote{{Temporal solitons and pulse compression in photonic crystal
  waveguides},} Nat Photon \textbf{4}, 862--868 (2010).

\bibitem{Yin:07}
L.~Yin and G.~P. Agrawal, \enquote{{Impact of two-photon absorption on
  self-phase modulation in silicon waveguides},} Opt. Lett. \textbf{32},
  2031--2033 (2007).

\bibitem{Monat:09}
C.~Monat, B.~Corcoran, M.~Ebnali-Heidari, C.~Grillet, B.~J. Eggleton, T.~P.
  White, L.~O'Faolain, and T.~F. Krauss, \enquote{{Slow light enhancement of
  nonlinear effects in silicon engineered photonic crystal waveguides},} Opt.
  Express \textbf{17}, 2944--2953 (2009).

\bibitem{Blanco-Redondo2014}
A.~Blanco-Redondo, C.~Husko, D.~Eades, Y.~Zhang, J.~Li, T.~F. Krauss, and B.~J.
  Eggleton, \enquote{{Observation of soliton compression in silicon photonic
  crystals},} Nat Commun \textbf{5}, 3160 (2014).

\bibitem{Rukhlenko:2010}
I.~D. Rukhlenko, M.~Premaratne, and G.~P. Agrawal, \enquote{{Nonlinear Silicon
  Photonics: Analytical Tools},} Selected Topics in Quantum Electronics, IEEE
  Journal of \textbf{16}, 200--215 (2010).

\bibitem{Gordon:86ssfs}
J.~P. Gordon, \enquote{{Theory of the soliton self-frequency shift},} Opt.
  Lett. \textbf{11}, 662--664 (1986).

\bibitem{Saleh:2011}
M.~F. Saleh, W.~Chang, P.~H\"{o}lzer, A.~Nazarkin, J.~C. Travers, N.~Y. Joly,
  P.~S.~J. Russell, and F.~Biancalana, \enquote{{Theory of
  Photoionization-Induced Blueshift of Ultrashort Solitons in Gas-Filled
  Hollow-Core Photonic Crystal Fibers},} Phys. Rev. Lett. \textbf{107}, 203902
  (2011).

\bibitem{Roy:13}
S.~Roy, A.~Marini, and F.~Biancalana, \enquote{{Self-frequency blueshift of
  dissipative solitons in silicon-based waveguides},} Phys. Rev. A \textbf{87},
  65803 (2013).

\bibitem{Perez:07}
V.~{P\'{e}rez Garc\'{\i}a}, P.~Torres, and G.~Montesinos, \enquote{{The Method
  of Moments for Nonlinear Schr\"{o}dinger Equations: Theory and
  Applications},} SIAM Journal on Applied Mathematics \textbf{67}, 990--1015
  (2007).

\bibitem{Gordon:86noise}
J.~P. Gordon and H.~A. Haus, \enquote{{Random walk of coherently amplified
  solitons in optical fiber transmission},} Opt. Lett. \textbf{11}, 665--667
  (1986).

\bibitem{Santhanam2003413}
J.~Santhanam and G.~P. Agrawal, \enquote{{Raman-induced spectral shifts in
  optical fibers: general theory based on the moment method},} Optics
  Communications \textbf{222}, 413--420 (2003).

\bibitem{Chen:10}
Z.~Chen, A.~J. Taylor, and A.~Efimov, \enquote{{Soliton dynamics in non-uniform
  fiber tapers: analytical description through an improved moment method},} J.
  Opt. Soc. Am. B \textbf{27}, 1022--1030 (2010).

\bibitem{Tsoy:06}
E.~N. Tsoy and C.~M. de~Sterke, \enquote{{Dynamics of ultrashort pulses near
  zero dispersion wavelength},} J. Opt. Soc. Am. B \textbf{23}, 2425--2433
  (2006).

\bibitem{Burgoyne:07}
B.~Burgoyne, N.~Godbout, and S.~Lacroix, \enquote{{Nonlinear pulse propagation
  in optical fibers using second order moments},} Optics Express \textbf{15},
  10075 (2007).

\bibitem{soref:87}
R.~A. Soref and B.~R. Bennett, \enquote{{Electrooptical effects in silicon},}
  Quantum Electronics, IEEE Journal of \textbf{23}, 123--129 (1987).

\bibitem{Boyraz:04}
O.~Boyraz, P.~Koonath, V.~Raghunathan, and B.~Jalali, \enquote{{All optical
  switching and continuum generation in silicon waveguides},} Opt. Express
  \textbf{12}, 4094--4102 (2004).

\bibitem{Corcoran:10}
B.~Corcoran, T.~D. Vo, M.~D. Pelusi, C.~Monat, D.-X. Xu, A.~Densmore, R.~Ma,
  S.~Janz, D.~J. Moss, and B.~J. Eggleton, \enquote{{Silicon nanowire based
  radio-frequency spectrum analyzer},} Opt. Express \textbf{18}, 20190--20200
  (2010).

\bibitem{Blanco:optica14}
A.~Blanco-Redondo, D.~Eades, J.~Li, S.~Lefrancois, T.~F. Krauss, B.~J.
  Eggleton, and C.~Husko, \enquote{{Controlling free-carrier temporal effects
  in silicon by dispersion engineering},} Accepted in Optica (2014).

\bibitem{Liao:13}
J.~Liao, M.~Marko, X.~Li, H.~Jia, J.~Liu, Y.~Tan, J.~Yang, Y.~Zhang, W.~Tang,
  M.~Yu, G.-Q. Lo, D.-L. Kwong, and C.~W. Wong, \enquote{{Cross-correlation
  frequency-resolved optical gating and dynamics of temporal solitons in
  silicon nanowire waveguides},} Opt. Lett. \textbf{38}, 4401--4404 (2013).

\bibitem{Wood:1991}
W.~M. Wood, C.~W. Siders, and M.~C. Downer, \enquote{{Measurement of
  femtosecond ionization dynamics of atmospheric density gases by spectral
  blueshifting},} Phys. Rev. Lett. \textbf{67}, 3523--3526 (1991).

\bibitem{Fedotov:2007}
A.~B. Fedotov, E.~E. Serebryannikov, and A.~M. Zheltikov,
  \enquote{{Ionization-induced blueshift of high-peak-power guided-wave
  ultrashort laser pulses in hollow-core photonic-crystal fibers},} Phys. Rev.
  A \textbf{76}, 53811 (2007).

\bibitem{Dorrer:02}
C.~Dorrer and I.~Kang, \enquote{{Simultaneous temporal characterization of
  telecommunication optical pulses and modulators by use of spectrograms},}
  Opt. Lett. \textbf{27}, 1315--1317 (2002).

\bibitem{bhat:2001}
N.~A.~R. Bhat and J.~E. Sipe, \enquote{{Optical pulse propagation in nonlinear
  photonic crystals},} Phys. Rev. E \textbf{64}, 56604 (2001).

\bibitem{bristow:07}
A.~D. Bristow, N.~Rotenberg, and H.~M. van Driel, \enquote{{Two-photon
  absorption and Kerr coefficients of silicon for 850-2200nm},} Applied Physics
  Letters \textbf{90}, 191104 (2007).

\end{thebibliography}
\end{document}